\documentclass[reprint]{revtex4-1}
\pdfoutput=1
\usepackage{amsmath}
\usepackage{graphicx}
\usepackage{enumerate}
\graphicspath{ {images/} }

\newcommand{\ket}[1]{| #1 \rangle }
\newcommand{\bra}[1]{\langle  #1 |}
\newcommand{\braket}[2]{\langle #1 | #2 \rangle}

\begin{document}
\title{Multiple-copy state discrimination of noisy qubits}
\author{Kieran Flatt }
\email{k.flatt.1@research.gla.ac.uk}
\author{Stephen M. Barnett}
\author{Sarah Croke}
\affiliation{School of Physics and Astronomy, University of Glasgow, Glasgow, G12 8QQ, UK}

\begin{abstract}
Multiple-copy state discrimination is a fundamental task in quantum information processing. If there are two, pure, non-orthogonal states then both local and collective schemes are known to reach the Helstrom bound, the maximum probability of successful discrimination allowed by quantum theory. For mixed states, it is known that only collective schemes can perform optimally, so it might be expected that these schemes are more resilient to preparation noise. We calculate the probability of success for two schemes, one local and one collective, in the regime of imperfect preparation fidelity. We find two surprising results. Firstly, both schemes converge upon the same many-copy limit, which is less than unity. Secondly, the local scheme performs better in all cases. This highlights the point that one should take into account noise when designing state discrimination schemes.
\end{abstract}

\maketitle

\section{Introduction}

In many quantum information processing tasks, one needs to identify, by measurement, the state of a system given that the finite and discrete set of states from which it is taken is known. This task is called state discrimination \cite{statediscriminationreview, baekwek, bergoureview, steveqi, chefles}. Unless the set of possible states is an orthogonal basis for some space they cannot be perfectly discriminated and instead the user usually seeks to minimise one of two figures of merit, either the probability of incorrectly identifying or failing to identify the state. The measurement which minimises the former of these is the Helstrom, or minimum-error, measurement \cite{helstrombook}. If there are two possible states, the optimal measurement has a simple analytic form. In more complex cases, such as three-or-more pure states \cite{haetal} or mixed states \cite{weirmixed, baehwang}, only limited results are known. 
\par
Our above comments relate to single-copy state discrimination. Given a resource of multiple systems, all prepared in the same state, it might be expected that the correlations can be used to improve the probability of success. This intuition is correct and the Helstrom bound, the optimal value of this probability, is known for discriminating two states. However, in this case a physical implementation of the measurement is typically hard to find. Furthermore, the issue of locality versus collectivity arises: can the bound be achieved with local measurements, those on individual systems, only or must the discriminator use collective measurements, which are more difficult to perform? It is known that the best measurement to discriminate multi-partite states is often a collective measurement, even for product states. Famous examples of this are the double trine ensemble \cite{pereswootters, massarpopescu, chitambarhsieh, crokeetaltrine} and the domino states \cite{bennettetal, childsetal, crokedomino}.
\par
For two-pure-state discrimination, it is known that a local scheme can reach the Helstrom bound \cite{acinetal, brodymeister, banetal}. In other scenarios, very few analytic results have been acquired and most knowledge comes from numerical simulations \cite{higginsetal, slussarenkoetal}. Here, some counterintuitive results emerge. One example is that the distinction between local and global optimality emerges. In some cases, among local schemes, the best overall measurement involves a fixed measurement on each qubit, which succeeds locally with a suboptimal probability \cite{higginsetal}. For a small number of copies, adaptive schemes perform better than fixed schemes \cite{higginsetal2}, but in the limit of large numbers of copies, this advantage disappears, even for mixed states \cite{higginsetal, higginsetal2, calsamigliaetal1, calsamigliaetal2, hayashi}. Further, it is for almost pure, but strictly speaking mixed, states that the gap in performance between collective stratgies and local strategies is most pronounced in the many-copy limit. Such unexpected results signal the need for further analytical work in this area. 
\par

The work presented in this paper investigates a separate, but related question. How resilient are multiple-copy state discrimination schemes to preparation noise? No real preparation is ever perfect, but for high enough fidelity we may consider the states to be pure. Further, decoherence properties of even state-of-the-art qubits can demonstrate significant variability (for a recent example see for examples Refs. \cite{burnettetal, schloretal}), resulting in a corresponding variability in the rate at which a preparation characterised as very high fidelity degrades over time. Finally, in a real-world physical communications system, instabilities in noise properties of a channel can lead to uncharacterised noise in the received states. How sensitive are schemes designed for pure states to a small amount of uncharacterized preparation noise? As the truly optimal scheme for noisy qubits will be collective, it might be expected that such schemes will be more resilient to preparation noise than the equivalent local scheme. Our approach is to compare two equivalent schemes, one local \cite{acinetal} and one collective \cite{quantumdatagathering}, both of which reach the Helstrom bound for discriminating two pure states. We apply each scheme, optimised for a specific pair of pure states, to the corresponding mixed states and relate the probability of success to the preparation fidelity. Our results show that, surprisingly, the local scheme consistently performs better than the collective scheme. Neither, however, approaches unit success probability as the number of copies, $N$, grows. Rather, they approach the same fixed bound. We discuss how to use information which would otherwise be thrown away in the local adaptive scheme to improve on this bound. This recovers asymptotic behaviour which, as the number of qubits approaches infinity, tends towards perfect discrimination.

\section{Preliminaries}
Two pure states of a qubit occupy a single great circle on the Bloch sphere. For this reason, they can be characterised in relation to each other by real numbers only and written in the form
\begin{equation} \label{cleanstate}
\ket{\psi_k} = \cos( \theta) \ket{0} + (-1)^k \sin( \theta) \ket{1} \: \: \: \: k=0,1.
\end{equation}
The overlap of these two states is $\braket{\psi_0}{\psi_1} = \cos(2 \theta)$ and, without loss of generality, $0 \leq \theta \leq \pi/4$. If a single system is prepared in either of these states with probabilities $p_k$, the highest possible probability of successful discrimination is given by the Helstrom bound,
\begin{equation}
{\rm P}^{H}_1 = \frac{1}{2} \left( 1 + \sqrt{ 1 - 4 p_0 p_1 \cos^2 ( 2 \theta) } \right).
\end{equation}
If $\theta = \pi/4$ the two states are orthogonal. In such a case, ${\rm P}^{H}_1 = 1$ and they can be perfectly discriminated. Otherwise, this quantity is less than one. The measurement which achieves this bound is a projective measurement onto the eigenvectors of $p_0 \ket{\psi_0} \bra{ \psi_0} - p_1 \ket{\psi_1} \bra{\psi_1}$. 
\par
If instead there is a resource of $N$ copies of the state, we are seeking to distinguish $\ket{\psi_0}^{\otimes N}$ from $\ket{\psi_1}^{\otimes N}$. As these can be considered as two single pure states on the total Hilbert space, the multiple-copy Helstrom bound is
\begin{equation} \label{helstrombound}
{\rm P}^{H}_N = \frac{1}{2} \left(1 + \sqrt{1 - 4p_0 p_1 \cos^{2N}(2\theta) } \right).
\end{equation}
In this case, the measurement which achieves this is again a von Neumann measurement, one that projects onto the eigenstates of $p_0 \ket{\psi_0} \bra{ \psi_0} ^ {\otimes N} - p_1 \ket{ \psi_1} \bra{\psi_1}^{\otimes N}$. To find these we must find the eigenvalues of a $2^N$ dimensional matrix, a task which is much simplified by the permutation symmetry in the many copy case. For pure states in particular, there are just two-dimensions that are important, and a number of optimal schemes are known, of which we consider two.
\par
In this article we are concerned with systems in which the resource qubits are prepared imperfectly. This is represented by a parameter $\delta \theta_i $ which characterises the displacement of the $i$th qubit's state from the ideal case such that
\begin{align} \label{noisystate}
\ket{\tilde{\psi}^{i}_k} &= \cos( \theta + \delta \theta_i) \ket{0} + (-1)^k \sin( \theta + \delta \theta_i) \ket{1} \nonumber \\
&= \cos(\delta \theta_i) \ket{\psi_k} - \sin( \delta \theta_i) \ket{\psi_{k \perp}}.
\end{align}
In the second line we have related the noisy form of the state to the ideal case, Eq. \ref{cleanstate} and introduce $\ket{\psi_{k \perp}}$ to indicate the state orthogonal to $\ket{\psi_k}$. The fidelity $F$ is the standard way to parameterise the noise on a system. It can be understood operationally as the probability that a measurement of the prepared state will identify it as the ideal state \cite{steveqi}. For pure states, it is defined as the overlap of the prepared and ideal states, averaged over the noise's probability distribution which we assume is symmetric, i.e., ${\rm P}( \delta \theta_i) = {\rm P}( - \delta \theta_i) $. One can consider this a Gaussian distribution however that level of detail is not required in what follows. The two noise parameters are then related by
\begin{equation}
\langle \cos^2 ( \delta \theta_i ) \rangle = \int | \braket{\tilde{\psi}^i_k}{\psi_k} |^2 {\rm P}(\delta \theta_i) = F ,
\end{equation}
where $1/2 \leq F \leq 1$. From this we also have
\begin{align}
\langle \sin^2 (\delta \theta_i) \rangle &= 1-F \\
\langle \cos(2 \delta \theta_i) \rangle &= 2F-1 \\
\langle \sin(2 \delta \theta_i ) \rangle &= 0.
\end{align}
The first two of these follows from the definition of the fidelity while the third uses the symmetry of the probability distribution. These are the only functions which are averaged in what follows. We assume that the noise on each qubit is independent of the others and average at each stage.
\par
Using these results we express the noisy form of the state, Eq. \ref{noisystate}, as a mixed state. We obtain
\begin{equation} \label{mixedstate}
\rho_k = F \ket{\psi_k} \bra{\psi_k} + (1-F) \ket{\psi_{k \perp}} \bra{\psi_{k \perp}},
\end{equation}
where we have averaged over the probabilty distribution of $\delta \theta_i$. If $F=1$ it is the relevant pure state. If instead $F=1/2$, which is the smallest possible value of the fidelity, it is a maximally mixed state, so that maximum noise erases all information about the state. For other values of $F$, the state varies monotically between these two points. Our interest throughout this paper will be in systems which are close to perfect fidelity.

\section{Local-adaptive measurement}
An important result in multiple-copy state discrimination is that it is possible to reach the Helstrom bound, Eq. \ref{helstrombound}, using local measurements only. We follow here the scheme of Ac{\'i}n \emph{et al} \cite{acinetal} but similar results have been found by others \cite{brodymeister, banetal}. They examine a local and adaptive scheme in which the measurement of the $n$th copy can depend upon the outcome of measurements on the previous $(n-1)$ copies. We first need to introduce some notation. The sequence of measurement outcomes is represented by a bit string $x$ as long as the number $N$ of qubits, with the $n$th result labelled $i_{n}$. The measurement onto the $n$th qubit is a projector onto the basis
\begin{align} \label{acinbasis}
\ket{\omega^n_0} &= \cos( \phi_x) \ket{0} + \sin( \phi_x) \ket{1} \nonumber \\
\ket{\omega^n_1} &= \sin( \phi_x ) \ket{0} - \cos( \phi_x) \ket{1}.
\end{align}
Here, we use $x$ for the bit string of the first $(n-1)$ results and adopt a different notation when it is required. In the local-adaptive measurement scheme, the measurement at each point depends on the previous outcomes in the scheme however the overall result is determined by the final measurement outcome alone. The optimal scheme of this kind turns out to be Bayesian updating \cite{acinetal}. On the first qubit, one projects onto the eigenvectors of $p_0 \ket{\psi_0} \bra{\psi_0} - p_1 \ket{\psi_1}\bra{\psi_1}$. On the rest, the relevant eigenbasis is instead ${\rm P}(0 |x) \ket{\psi_0} \bra{\psi_0} - {\rm P}(1|x) \ket{\psi_1}\bra{\psi_1}$, in which ${\rm P}(k|x)$ is the probability, calculated from Bayes' theorem, that the state $ \ket{\psi_k}$ was prepared given that bit string $x$ is the measurement record. The $\phi_x$ for which this measurement satisifies the Helstrom bound is found to be 
\begin{equation} \label{locadopt}
\cos( 2 \phi_x ) = (-1)^{i_{N-1}} \sqrt{ \frac{ 1 - 4 p_0 p_1 \cos^{2N-2}(2 \theta) }{1- 4 p_0 p_1 \cos^{2N}(2 \theta) } }.
\end{equation}
The only appearance of the bit string $x$ here is in the single index $i_{N-1}$, which is the value of the prior measurement. Thus the scheme does not use the entire measurement and is in this sense Markovian as well as Bayesian.
\par
Here we apply the local-adaptive scheme, in the form optimised for pure states, to the mixed states relevant to imperfect preparation. A true Bayesian scheme, one that uses the entire measurement record, would be the best way to generalise the scheme to mixed states. We return to this point later. For now, we are interested in a direct comparison of the pure state schemes and so proceed with the Markovian form.
\par
We begin by showing that this scheme reaches the Helstrom bound in the case of perfect preparation. We use a different approach to that in Ref. \cite{acinetal} as it does not generalise straightforwardly to include noise. This calculation gives a form for the success probability with $N$ qubits in terms of that for $(N-1)$ qubits, an inductive formula which is solved by the Helstrom bound. We then modify the calculation to include noise. This leads to a different inductive formula, which is then solved to give the overall success probability. In thse calculations, we make repeated use of the result
\begin{align} \label{locadprob1}
{\rm P}(i_{N} | x, k ) = \frac{1}{2} \left[ 1 \right. &+ (-1)^{i_N} \cos(2\theta) \cos(2 \phi_x) \\
& \left. + (-1)^{i_N + k} \sin(2\theta) \sin(2 \phi_x) \right] \nonumber
\end{align}
for the probability that the $N$th outcome is $i_N$ given that the state $\ket{\psi_k}$ was sent and that the initial $(N-1)$ results were $x$. This is calculated using Eqs. \ref{cleanstate} and \ref{acinbasis}. 
\par
In the local-adaptive scheme, the identification of the prepared state is made with the final outcome. For this reason, the probability of success is
\begin{equation}
{\rm P}^{ad}_N = \sum_{x, k} p_k {\rm P}( k | x, k ) {\rm P}( x | k ) .
\end{equation}
This is a sum over both signal states $k=0,1$ and over all bit strings $x$ of length $(N-1)$, none of which contribute directly to the state identification. We first substitute Eq. \ref{locadprob1}, with $i_{N} = k$, into this result to give
\begin{align} \label{locadprob2}
{\rm P}^{ad}_N = \frac{1}{2} \left[ 1 \right. &+  \sum_{x,k} \left(  \sin(2\theta) \sin(2\phi_x)  p_k {\rm P}(x|k) \right.   \\
& \left. \left. +  \cos(2\theta) \cos(2 \phi_x ) (-1)^{i_{N-1}+k} p_k {\rm P}( x | k) \right) \right]. \nonumber
\end{align}
Then next step is to use Eq. \ref{locadopt} for the optimal value of $2 \phi_x$ in this equation:
\begin{align} \label{adcalc1}
& {\rm P}^{ad}_N = \frac{1}{2} \left[ 1+  \frac{ \sin^2 (2 \theta) }{\sqrt{1-\cos^{2N}(2 \theta)}} \sum_{x,k} p_k {\rm P}(x|k)  \right. \\
&+ \left. \cos^2 ( 2 \theta) \sqrt{ \frac{1 -\cos^{2N-2}(2 \theta)}{1 - \cos^{2N}(2\theta)}} \sum_{x,k} (-1)^{i_{N-1}+k} p_k {\rm P}( x | k) \right]. \nonumber
\end{align}
The first sum in this expression is straightforward to evalulate. It is simply a sum over a complete set of possible scenarios and we have $\sum_{x,k} p_k {\rm P}(x | k) = 1$. The other series is a little more complicated. We use the usual rules of conditional probability to write
\begin{equation}
{\rm P}(x | k) = {\rm P}( i_{N-1} \dot{x} | k ) = {\rm P}( i_{N-1} | \dot{x}, k) {\rm P}(\dot{x} | k),
\end{equation}
where we introduce the notation $\dot{x}$ for the bit string of the first $(N-2)$ results. We use also Eq. \ref{locadprob1}, with $x$ replaced by $\dot{x}$ and $i_{N}$ replaced by $i_{N-1}$, for the probabilities ${\rm P}( i_{N-1} | \dot{x}, k)$ in this equation. Bringing together all of these results, a short calculation reveals
\begin{align}
&\sum_{x,k} (-1)^{i_{N-1} + k } p_k {\rm P}(x | k ) \\
&=  \sum_{\dot{x},k} \left( \sin(2\theta) \sin(2\phi_{\dot{x}})  p_k {\rm P}(\dot{x}|k)  \right. \nonumber \\
&\left. +  \cos(2\theta) \cos(2 \phi_{\dot{x}} ) (-1)^{i_{N-2}+k} p_k {\rm P}( \dot{x} | k) \right). \nonumber
\end{align}
This should be compared with Eq. \ref{locadprob2}, in which the same expression occurs but over the final rather than penultimate outcome. This can be used to write the expression as
\begin{equation}
\sum_{x,k} (-1)^{i_{N-1} + k } p_k {\rm P}(x | k ) = 2 P^{ad}_{N-1} -1.
\end{equation}
After subsituting this into Eq. \ref{adcalc1}, we are left with the inductive expression 
\begin{align} \label{noiselessprob}
{\rm P}^{ad}_{N} = \frac{1}{2} \left[ \vphantom{\frac{\sqrt{1}}{12}} 1 \right. &+ \frac{ \sin^2 (2 \theta)}{\sqrt{1 - \cos^{2N}(2 \theta)}} \\
& \left. + \cos^2 (2 \theta) \sqrt{ \frac{1 - \cos^{2N-2}(2 \theta) }{1 - \cos^{2N}(2 \theta)}} (2 {\rm P}^{ad}_{N-1} - 1) \right]. \nonumber
\end{align}
The general solution to this equation is the multiple-copy Helstrom bound, Eq. \ref{helstrombound}, which can be verified by direct substitution. The $N=1$ case corresponds to single-copy state discrimination and that bound is derived in the usual manner. That the probability expression has this inductive form follows as the measurement strategy is Markovian. We have followed others in showing that the Helstrom bound can be reached with local measurements only \cite{acinetal, brodymeister, banetal}. Our main result in this section is a generalisation of this expression to the regime of imperfect preparation fidelity. 
\par
The calculation proceeds in the same manner as that without noise. The difference is in the probability of a specific result $i_{N}$ given that the state $\ket{\psi_k}$ was prepared, which changes when the latter is replaced with a noisy state. To take this into account, Eq. \ref{locadprob1} is replaced by an equivalent expression calculated using Eqs. \ref{mixedstate} and \ref{acinbasis}. The new probability is 
\begin{align}
& {\rm P}(i_{N} | x, k )  \nonumber \\
&= \frac{1}{2} \left[ 1 + (2F - 1) (-1)^{i_N} \cos(2 \theta) \cos(2 \phi_x) \right. \\
& \left. \hspace{1cm} +(2F-1) (-1)^{i_N + k} \sin(2 \theta) \sin(2 \phi_x)  \right] \nonumber
\end{align}
so that the only change in the noisy case is the appearance of the factor $(2F-1)$ here. We use this to derive, in exactly the same manner as in the perfect-fidelity case, the probability of success. The result of this, as might be expected based on the change to the individual probabilities, is simply
\begin{align}
{\rm P}^{ad}_{N} 
&= \frac{1}{2} \left[ 1 + (2F-1) \frac{ \sin^2 (2 \theta)}{\sqrt{1 - \cos^{2N}(2 \theta)}}  \right. \\
& \left.+ (2F-1) \cos^2 (2 \theta) \sqrt{ \frac{1 - \cos^{2N-2}(2 \theta) }{1 - \cos^{2N}(2 \theta)}} ( 2 {\rm P}^{ad}_{N-1} - 1) \right]. \nonumber
\end{align}
This relation is hardly more complicated than the noisless case, Eq. \ref{noiselessprob}, but its solution is much more complicated. By recursive application of this formula using the $N=1$ case, which can be evaluated analytically, we establish that the solution is
\begin{align} \label{locadsol1}
{\rm P}^{ad}_N = \frac{1}{2} \left( \vphantom{\frac{x^2}{2}} 1 \right. & + (2F-1)^N \sqrt{1-\cos^{2N}(2 \theta) } \\
& \left. + \frac{\sin^2(2 \theta) }{\sqrt{1 - \cos^{2N}(2 \theta) } } \mathcal{S}_N \right), \nonumber
\end{align}
where we introduce the notation
\begin{equation}
\mathcal{S}_N = \sum^{N}_{i=1} (2F-1)^{N+1-i} (1 - (2F-1)^{i-1} ) \cos^{2N-2i}(2 \theta). \nonumber
\end{equation}
This solution can be verified by substitution into the inductive relationship. The series $\mathcal{S}_N$ can be evaluated using the usual formulae for geometric progressions. After some algebraic manipulation we find
\begin{align} \label{locadsol2}
\mathcal{S}_N = &(2F-1) \frac{ 1 - (2F-1)^N \cos^{2N}(2\theta) }{1 - (2F-1)\cos^2 ( 2\theta) }\nonumber \\
&- (2F-1)^N \frac{1 - \cos^{2N}(2 \theta)}{1 -\cos^2 (2 \theta)}.
\end{align}
Between Eqs. \ref{locadsol1} and \ref{locadsol2}, the probability that the local-adpative scheme successfully identifies the state is defined in terms of the preparation fidelity $F$. In the perfect-fidelity case $F=1$, substitution shows that $\mathcal{S}_N = 0$, and we have that the usual Helstrom bound is achieved. If instead $F=1/2$, the prepared state is by definition a completely mixed state for both $\ket{\psi_0}$ and $\ket{\psi_1}$, so that the states are indistinguishable. For this value of the fidelity, the probability becomes one-half which corresponds to guessing. The other interesting limit is the behaviour of the scheme if there are many copies of the state. We look at this in a later section where we also plot the success probability. 

\section{Quantum data gathering}
The previous measurement scheme is purely local. It produces a classical bit value for each of the resource qubits. It is known that schemes of this type are in general not able to perform optimal state discrimination when the possible states are mixed. One requires collective measurements. Here, we are interested in the ability of these schemes to function in the presence of preparation noise. As an example of one scheme which measures collectively, we consider quantum data gathering \cite{quantumdatagathering}.
\par
This scheme requires a quantum memory, a qubit which does not decohere between interactions. This probe is initialised in the state $\ket{0}$. When required, we label this space $\mathcal{H}_A$. The interaction with the first qubit is a SWAP gate. The remaining interactions leave the resource qubits, labelled $S_i$ (where $i = 1,2,...,N$), each in the state $\ket{0}$ and, if there is no preparation noise, leave the probe in one of two states
\begin{equation}
\ket{\psi^{(n)}_k} = \cos(\theta_n) \ket{0} + (-1)^k \sin(\theta_n) \ket{1}, \: \: \: k=0,1
\end{equation}
in which
\begin{equation} \label{thetan}
\cos(\theta_n) = \sqrt{ \frac{1}{2} \left( 1 + \cos^n(2 \theta) \right) } .
\end{equation}
These two states have an overlap $\braket{\psi^{(n)}_0}{\psi^{(n)}_1} = \cos^n (2 \theta)$, where $n$ is the number of qubits which the probe has interacted with until that point in the scheme. Thus, the probability of success is the Helstrom bound. The protocol works as the product state of the $N$ systems exists in a two-dimensional subspace of the overall Hilbert space. This, of course, no longer holds for mixed states, for which some information about the states will be retained in the resource qubits. The interactions between the probe and resource qubits is a unitary operation which maps this subspace onto the two dimensions of the probe's space through the index $k$, which is the only piece of information needed to characterise each state. The unitary operator $U_n$ that performs such an operation has the property $U_n \ket{\psi_k}_{S_n} \ket{\psi^{(n-1)}_k}_A = \ket{0}_{S_n} \ket{\psi^{(n)}_k}_A$. Alone, this does not span the Hilbert space and we need to include also the state's components which appear only if the preparation is imperfect. The choice we make is $U_n \ket{\psi_{k \perp}}_{S_n} \ket{\psi^{(n-1)}_{k \perp}}_A = \ket{1}_{S_n} \ket{\psi^{(n)}_{k \perp}} $. The unitary operator, written in the computational basis for both qubits, is
\begin{align} \label{qdgunitary}
U_n \ket{0_{S_n} 0_{A} } &= \frac{ \cos(\theta) \cos(\theta_{n-1})}{\cos(\theta_n)} \ket{0_{S_n} 0_A} \nonumber \\
& \hspace{1cm} + \frac{ \sin(\theta) \sin(\theta_{n-1})}{\cos(\theta_n)} \ket{1_{S_n} 0_A} \nonumber \\
U_n \ket{1_{S_n} 1_{A} } &= \frac{ \sin(\theta) \sin(\theta_{n-1})}{\cos(\theta_n)} \ket{0_{S_n} 0_A} \nonumber \\
&\hspace{1cm}- \frac{ \cos(\theta) \cos(\theta_{n-1})}{\cos(\theta_n)} \ket{1_{S_n} 0_A} \nonumber \\
U_n \ket{1_{S_n} 0_A} &= \frac{\sin(\theta) \cos(\theta_{n-1})}{\sin ( \theta_{n})} \ket{0_{S_n} 1_A} \nonumber \\
&\hspace{1cm}+ \frac{ \cos(\theta) \sin(\theta_{n-1})}{\sin(\theta_n)} \ket{1_{S_n} 1_A} \nonumber \\
U_n \ket{ 0 _{S_n} 1_A} &= \frac{ \cos(\theta) \sin( \theta_{n-1})}{\sin(\theta_n)} \ket{0_{S_n} 1_A} \nonumber \\
&\hspace{1cm}- \frac{ \sin(\theta) \cos(\theta_{n-1})}{\sin(\theta_n)} \ket{1_{S_n} 0_A}.
\end{align}
After all resource qubits have been processed, the qubit is measured with a Helstrom measurement which corresponds to distinguishing $\ket{\psi^{(N)}_0}$ from $\ket{\psi^{(N)}_1}$. Again, the quantity we calculate is the probability that this measurement is successful if the prepared qubits are instead mixed states. 
\par
The strategy that we use to calculate this probability is to find, by representing the interactions as Kraus operators acting on $\mathcal{H}_A$, the probe's state at each stage of the protocol. These Kraus operators are derived by considering that the resource qubits are subsequently measured in the computational basis though we sum over both outcomes. This strategy gives us the possibility of considering that such a measurement, which could be used as a diagnostic for the protocol's behaviour, does occur. The Kraus operators are calculated as $M^{(n)}_{i,k} = \bra{i}_{S_n} U_n \ket{\tilde{\psi}^{n}_k}_{S_n}$, where $i=0,1$ and we use the noisy form of the state. As the calculation involves pairs of Kraus operators, at this point we do not average over the noise. The Kraus operators are best expressed in the form
\begin{align}
M^{(n)}_{0,k} \ket{\psi^{(n-1)}_k} &= \cos( \delta \theta_n) \ket{ \psi^{(n)}_k}  \\
M^{(n)}_{0,k} \ket{\psi^{(n-1)}_{k \perp}} &= -\frac{ \sin(2 \theta) \sin( \delta \theta_n) \cos(2 \theta_{n-1})}{\sin(2 \theta_n)} \ket{ \psi^{(n)}_k} \nonumber \\
&+ \frac{ \cos(2 \theta + \delta \theta_n) \sin(2 \theta_{n-1})}{\sin(2 \theta_n)} \ket{\psi^{(n)}_{k \perp}} \nonumber \\
M^{(n)}_{1,k} \ket{\psi^{(n-1)}_k} &= \sin( 2 \theta + \delta \theta_n) \ket{ \psi^{(n)}_{k \perp}} \nonumber \\
M^{(n)}_{1,k} \ket{ \psi^{(n-1)}_{k \perp}} &= - \frac{ \sin( \delta \theta_n) \sin( 2 \theta_{n-1}) }{ \sin(2 \theta_n)} \ket{ \psi^{(n)}_{k}} \nonumber \\
& \hspace{-2cm}+ \frac{ \sin( \delta \theta_n) \cos( 2 \theta_{n-1}) - \sin( 2 \theta + \delta \theta_n) \cos( 2 \theta_n)}{\sin(2 \theta_n)} \ket{\psi^{(n)}_{k \perp}}. \nonumber
\end{align}
One way to think about these is objects is that the outcome $M^{(n)}_{0,k}$ indicates that the protocol is running well and conversely for $M^{(n)}_{1,k}$. This is because the former is the only outcome if the fidelity is perfect. This is seen in the Kraus representation as the action of $M^{(n)}_{0,k}$ is to map the state $\ket{\psi^{(n-1)}_k}$ onto $\ket{\psi^{(n)}_k}$, thus preserving the information which is encoded in that basis, whereas the operator $M^{(n)}_{1,k}$ has the opposite effect: by mapping the state $\ket{\psi^{(n-1)}_k}$ onto $\ket{\psi^{(n)}_{k\perp}}$ it deletes all the information which has been acquired up to that point. This is the origin of the claim that a subsequent measurement of the prepared qubit can act as a diagnostic. This point is later considered in more detail. 
\par
We are now in a position to calculate the density matrix of the probe after $N$ interactions. We assume that all noise is in the state preparation and that the operations are implemented perfectly. At the first step, the sample is swapped with the probe, so that the probe is left in the state
\begin{equation}
\rho_{1} = F \ket{\psi_k}\bra{\psi_k} + (1-F) \ket{\psi_{k \perp}} \bra{\psi_{k \perp}},
\end{equation}
as was shown earlier (to simplify the notation, we drop the index $k$ from the density operator $\rho$). We evaluate the next step in full detail and the result allows us to find, by inspection, the form of the density matrix in general. We multiply by the Kraus operators and then average over $\delta \theta_i$ in one step here. It is a straightforward (though longwinded) calculation to find
\begin{align}
\rho_2 &= M^{(2)}_{0,k} \rho_1 M^{(2) \dagger}_{0,k} + M^{(2)}_{1,k} \rho_1 M^{(2) \dagger}_{1,k} \nonumber \\
&= \left( F - (1-F)(2F-1) \cos^2(2 \theta) \right) \ket{\psi^{(2)}_k}\bra{\psi^{(2)}_k} \nonumber \\
&+ \left (1 - F + (1-F)(2F-1) \cos^2 (2 \theta) \right) \ket{ \psi^{(2)}_{k \perp}} \bra{ \psi^{(2)}_{k \perp}} \nonumber \\
& + (1-F) \frac{ \cos^2(2 \theta) \sin^2(2 \theta)}{ \sin^2 (2 \theta_2)} \sigma^{(2)}_x,
\end{align}
where $\sigma^{(2)}_x = \ket{\psi^{(2)}_k} \bra{ \psi^{(2)}_{k \perp}} +  \ket{\psi^{(2)}_{k \perp}} \bra{ \psi^{(2)}_{k}}$. We keep the convention of using a superscript on the Pauli matrix to indicate the basis in which it is written. This density matrix can be understood as two pieces: a trace-one, diagonal piece consisting of the first two terms and another consiting of only the $\sigma_x$ matrix. We can expect, based on this, that the same is true of the general density matrix, which we expect can be written
\begin{align} \label{qdgrhoguess}
\rho_{N} &= A_N \ket{ \psi^{(N)}_k} \bra{ \psi^{(N)}_k} + (1- A_N) \ket{ \psi^{(N)}_{k \perp}} \bra{ \psi^{(N)}_{k \perp}} \nonumber \\
&+ B_N \sigma^{(N)}_x .
\end{align}
This is confirmed by the following analysis, in which we evaluate $A_N$ and $B_N$ by calculating how each piece (diagonal and Pauli) is updated. We again multiply by the Kraus operators and average over $\delta \theta_i$ in a single step. The first result is
\begin{align}
&A_{N-1} \sum_i M^{(n)}_{i,k}  \ket{\psi^{(N-1)}_k} \bra{ \psi^{(N-1)}_k} M^{(n) \dagger}_{i,k} \nonumber \\
&+  (1 - A_{N-1})\sum_i M^{(n)}_{i,k}   \ket{\psi^{(N-1)}_{k \perp}} \bra{ \psi^{(N-1)}_{k \perp}} M^{(n) \dagger}_{i,k} \nonumber \\
&= \left( F - (1-A_{N-1}) (2F-1)\cos^2 (2 \theta) \right) \ket{ \psi^{(N)}_k} \bra{ \psi^{(N)}_k} \nonumber \\
&+ \left( 1 - F +(1 - A_{N-1}) (2F-1) \cos^2 ( 2 \theta) \right) \ket{ \psi^{(N)}_{k \perp}} \bra{ \psi^{(N)}_{k \perp}} \nonumber \\
& - \left(1 - A_{N-1} \right) \frac{ (2F-1) \sin^2(2 \theta) \cos(2 \theta_N)}{\sin(2 \theta_N)} \sigma^{(N)}_x.
\end{align}
Notice that again we find the same structure, that of a diagonal piece and a Pauli matrix. The other update is 
\begin{align}
&\sum_i M^{(n)}_{i,k} \sigma^{(N-1)}_{x} M^{(n) \dagger}_{i,k} \nonumber \\
& \hspace{1cm}= \frac{ (2F-1) \cos(2 \theta) \sin(2 \theta_{N-1})}{ \sin(2 \theta_N)} \sigma^{(N)}_x.
\end{align}
It is seen that both terms contribute in the form, written in the natural basis of the next step, that we have predicted and the density matrix will always take the form of Eq. \ref{qdgrhoguess}. Repeated application of the above two results allow us to evaluate $A_N$ and $B_N$, which are both written in terms of geometric progressions. We find 
\begin{align}
A_N &= F \cos^{2N-2}(2 \theta) (2F-1)^{N-1}  \\
&+ \left( F - (2F-1) \cos^2(2 \theta) \right) \sum^{N-2}_{i=0} \cos^{2i}(2 \theta) (2F-1)^i \nonumber \\
B_N &= (2F-1) \sin^2 ( 2 \theta)  \\
& \hspace{0.4cm} \times \sum^{N-1}_{i=1} \frac{ \cos(2 \theta_{i+1})}{\sin(2 \theta_{i+1})} \prod^{N}_{j = i+2} (2F-1) \cos(2 \theta) \frac{\sin(2 \theta_{j-1})}{\sin(2 \theta_j)}. \nonumber
\end{align}
It is straightforward to evaluate the summation to give
\begin{equation} \label{ansol}
A_N = 1 - (1-F) \frac{ 1 - \cos^{2N}(2\theta) (2F-1)^N }{1 - \cos^2 ( 2 \theta) (2F-1)},
\end{equation}
which is then used alongside Eq. \ref{thetan} to evaluate
\begin{align} \label{bnsol}
&B_N = (1-F) \frac{ \sin^2 ( 2 \theta) \cos^{N-1}(2\theta)}{\sin(2 \theta_N)}   \\
& \times \left[ \frac{1 - (2F-1)^{N-1}}{1-(2F-1)} \right. \nonumber \\
& \left. \hspace{1cm} - \cos^{N+1}(2 \theta) (2F-1)^{N-1} \frac{1 - \cos^{2N-2}(2 \theta)}{1- \cos^2 (2 \theta)} \right]. \nonumber 
\end{align}
The denominators of the two fractions inside the braces could each be simplified however we leave them in this form so that it is clear that there are no convergence issues in the limit $F \rightarrow 1/2$ or $\theta \rightarrow 0$. Between Eqs. \ref{qdgrhoguess}, \ref{ansol} and \ref{bnsol}, we have characterised the probe's density matrix in terms of the fidelity and state parameters only. If $F=1$, which corresponds to the perfect fidelity case,  $A_N=1$ and $B_N=0$. This corresponds to the probe being in the pure state $\ket{\psi^{(N)}_k}$, as one would expect. If instead $F=1/2$, which corresponds to maximum infidelity, then again $B_N = 0$ however here $A_N=1/2$. This means that the probe is in a maximally mixed state so that it carries no information about the prepared state. This corroborates with the analysis of the similar cases in the local-adaptive scheme.
\par
In the quantum data gathering routine, following the unitary interactions between the probe and all resource qubits, the probe is left in the density matrix that we have calculated. If the fidelity is perfect, this will be one of two possible states, either $\ket{\psi^{(N)}_0}$ or $\ket{\psi^{(N)}_1}$. At this stage in the protocol, the probe is then measured with the Helstrom measurement which best distinguishes these states. This is the final piece of the calculation, which gives us the probability ${\rm P}^{qdg}_{N}$ that the scheme succeeds. Helstrom's conditions tell us that the best measurement is a projector onto the eigenvalues of $p_0 \ket{\psi^{(N)}_0} \bra{\psi^{(N)}_0} - p_1 \ket{\psi^{(N)}_1} \bra{\psi^{(N)}_1}$. The case $p_0 \neq p_1$ is significantly more involved without adding further understanding. For this reason we restrict our attention to equiprobable preparation $p_0 = p_1 = 1/2$ for this scheme. The relevant eigenvectors are
\begin{align}
\ket{\psi^{(N)}_+} &= \sqrt{ \frac{1 + \sin(2 \theta_N)}{2}} \ket{ \psi^{(N)}_k} + \sqrt{ \frac{1- \sin(2 \theta_N)}{2}} \ket{ \psi^{(N)}_{k \perp}} \nonumber \\
\ket{\psi^{(N)}_-} &= \sqrt{ \frac{1- \sin(2 \theta_N)}{2}} \ket{ \psi^{(N)}_k} - \sqrt{ \frac{1 + \sin(2 \theta_N)}{2}} \ket{ \psi^{(N)}_{k \perp}},
\end{align}
where the subscript $\pm$ indicates an associated eigenvalue of $\lambda = \pm 1$. It is the uppermost of these which corresponds to the correct outcome. The success probability derived from this
\begin{align} \label{qdgprob}
{\rm P}^{qdg}_{N} &= \bra{\psi^{(N)}_+} \rho_N \ket{\psi^{(N)}_+}  \\
&= \frac{1 - \sin(2 \theta_N)}{2} + A_N \sin(2 \theta_N) - B_N \cos(2 \theta_N). \nonumber
\end{align}
The Helstrom bound is written in a form useful here as ${\rm P}^{H}_N = ( 1 + \sin(2 \theta_N))/2$. We see that ${\rm P}^{qdg}_N$ is leading order in the Helstrom bound (once $A_N$ and $B_N$ are entered), followed by terms which are linearly and inversely proportional to that object. This structure is similar to the equivalent expression for the local-adaptive measurement scheme. Eqs. \ref{ansol}, \ref{bnsol} and \ref{qdgprob} together define the probability of success for the quantum data gathering.

\section{Discussion}

\begin{figure*}
  \begin{tabular}{c@{\hskip 2cm}c}
  \includegraphics[width=7cm]{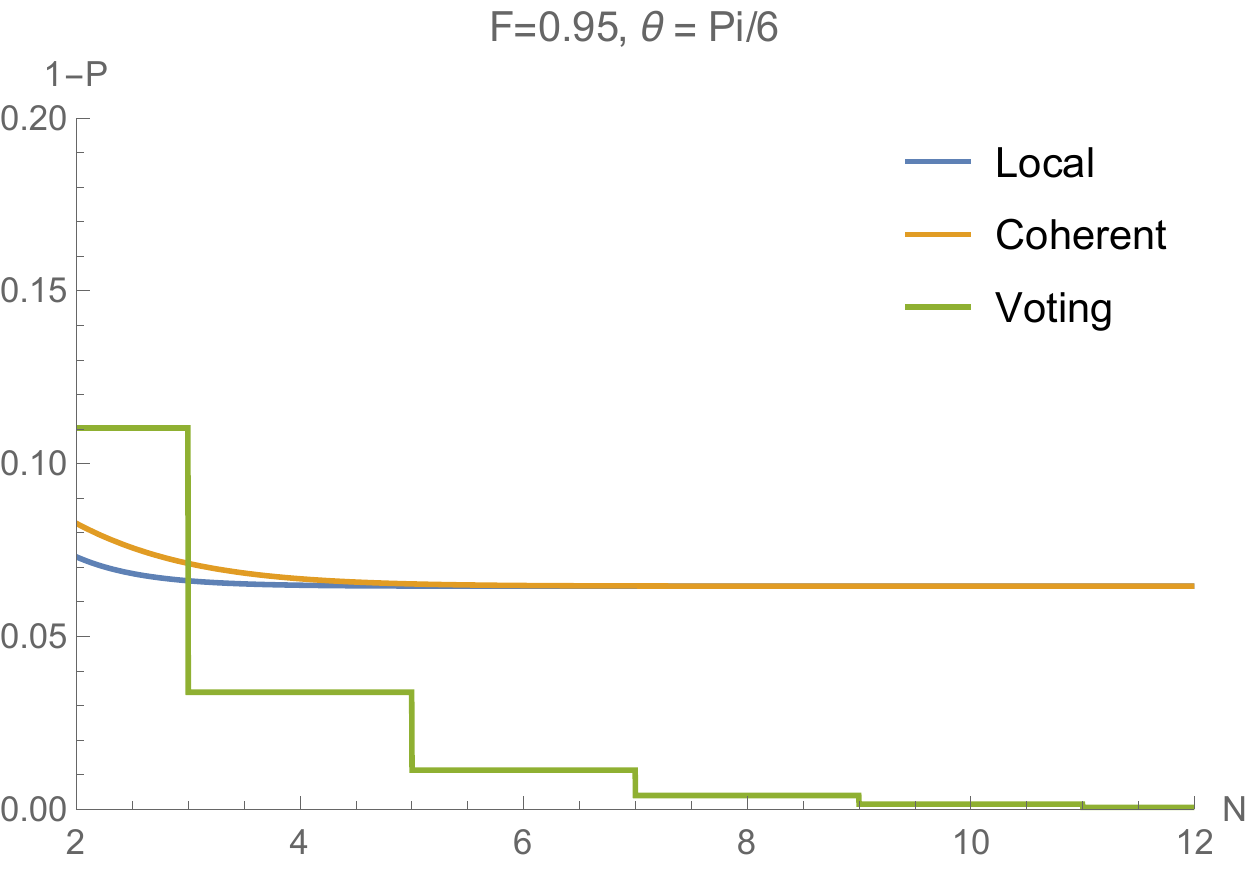} &
  \includegraphics[width=7cm]{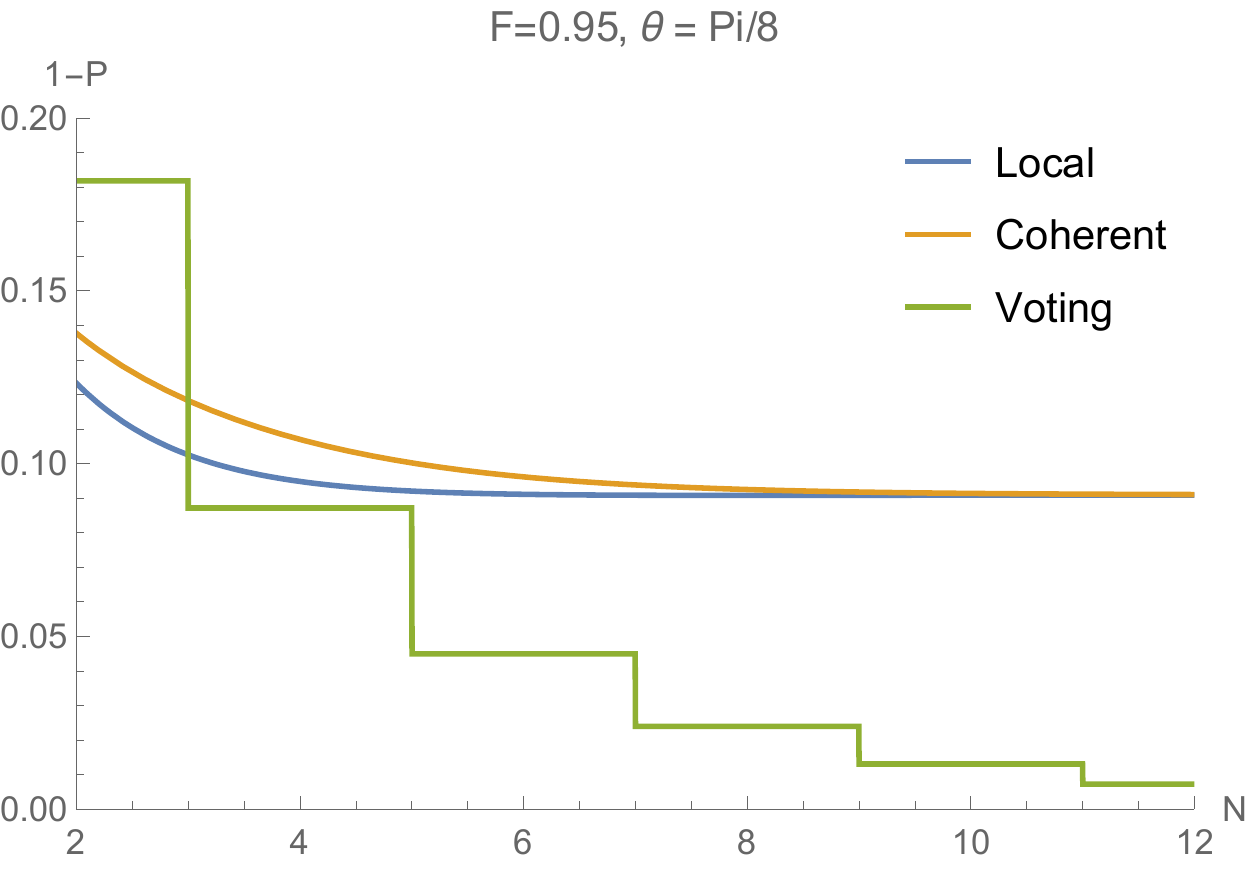} \\
  \vphantom{\includegraphics[height=1cm]{f095tpi6_new} }& \\
  \includegraphics[width=7cm]{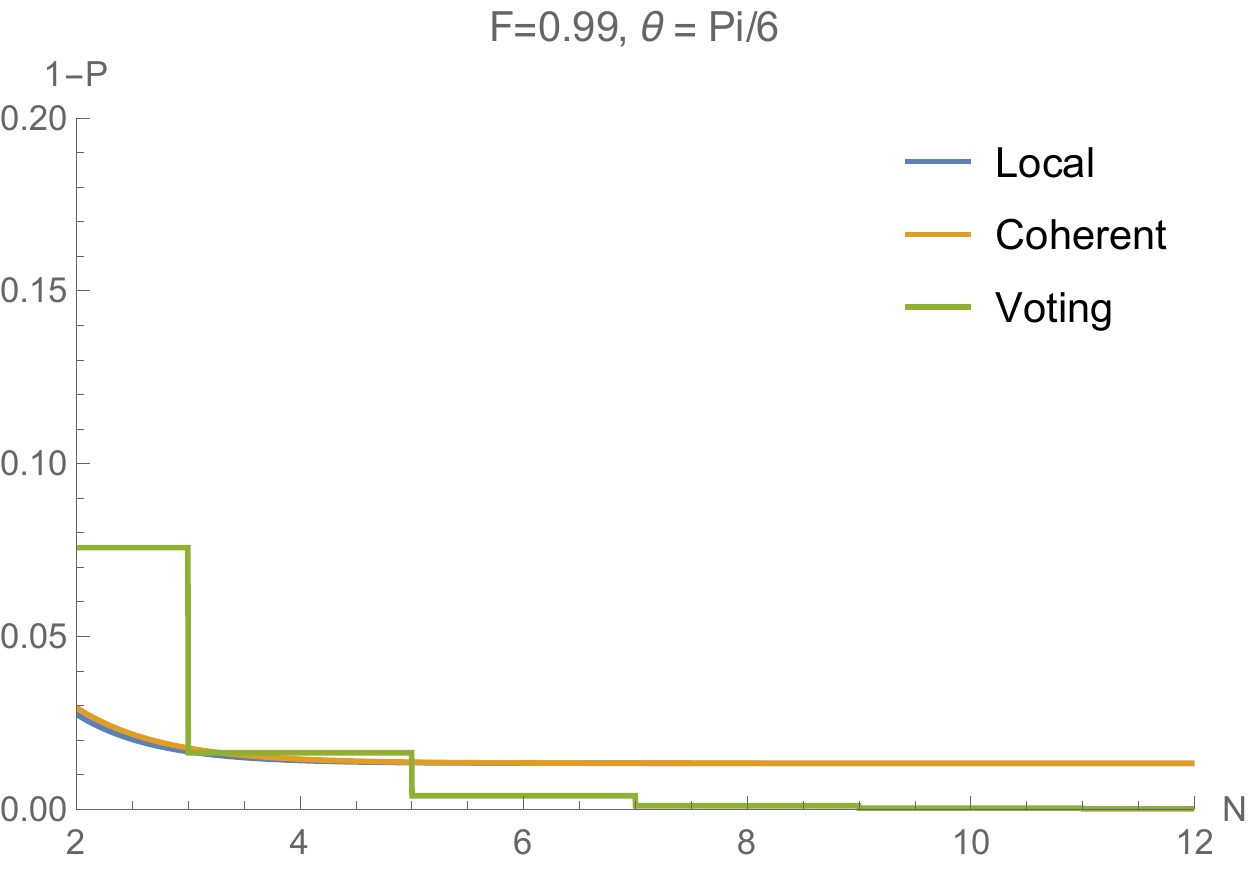} &
  \includegraphics[width=7cm]{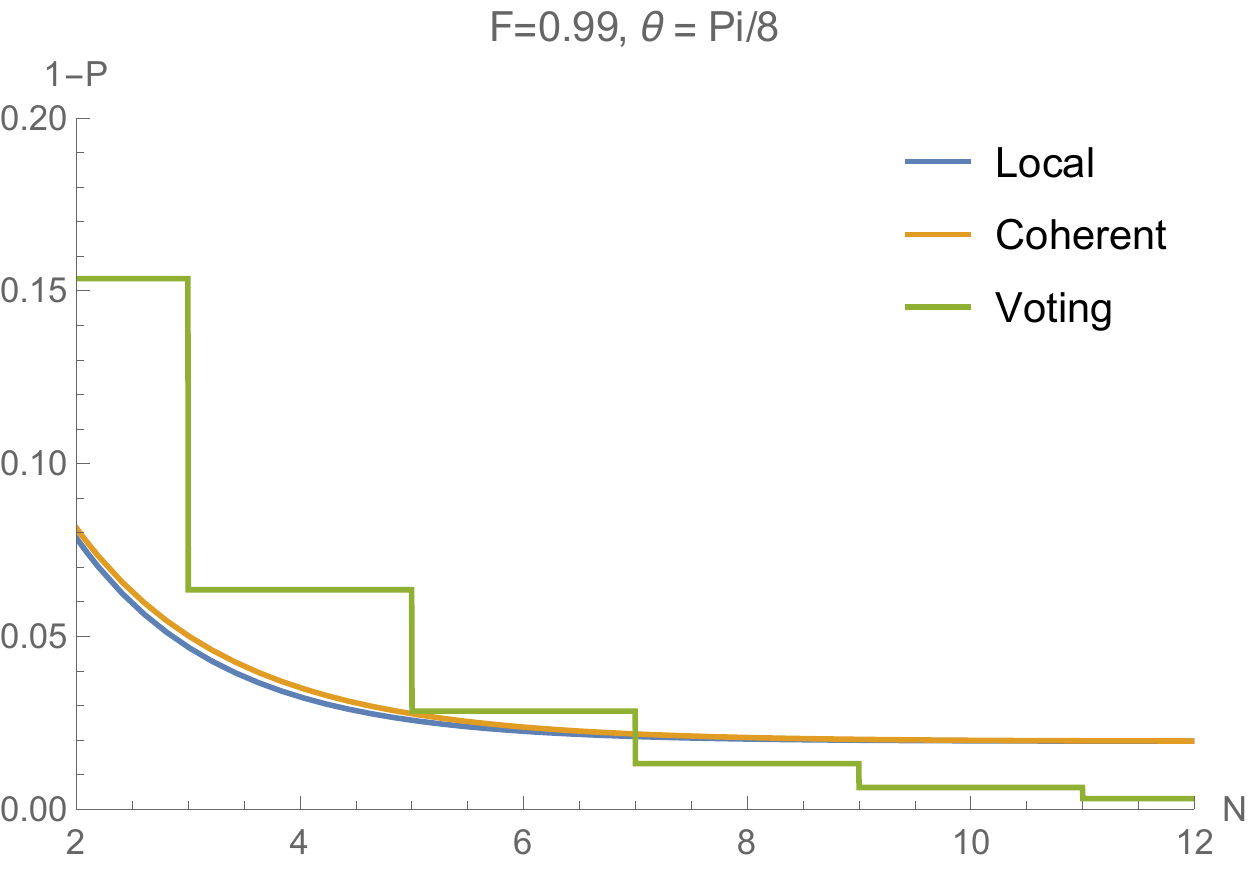} \\
  \vphantom{\includegraphics[height=1cm]{f095tpi6_new} }& \\
  \includegraphics[width=7cm]{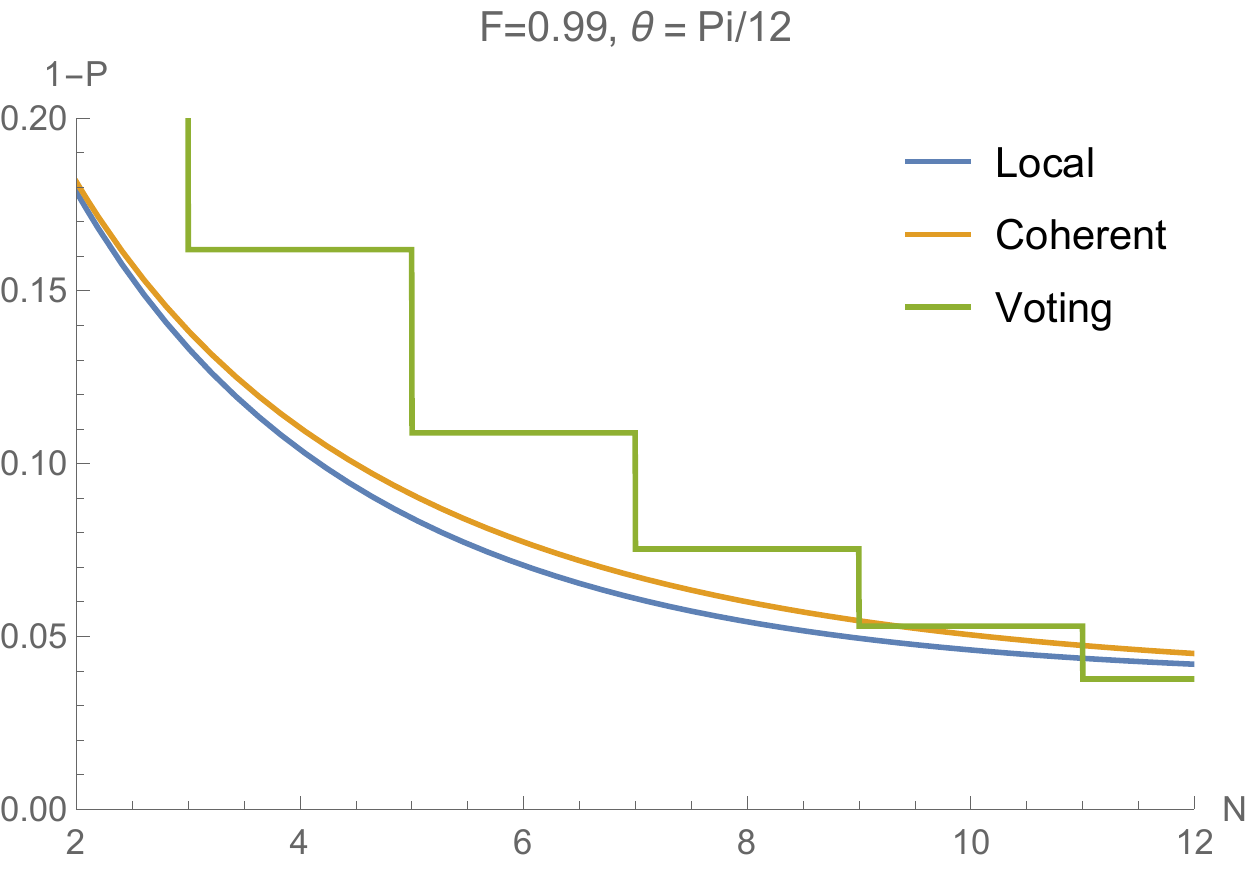} &
  \includegraphics[width=7cm]{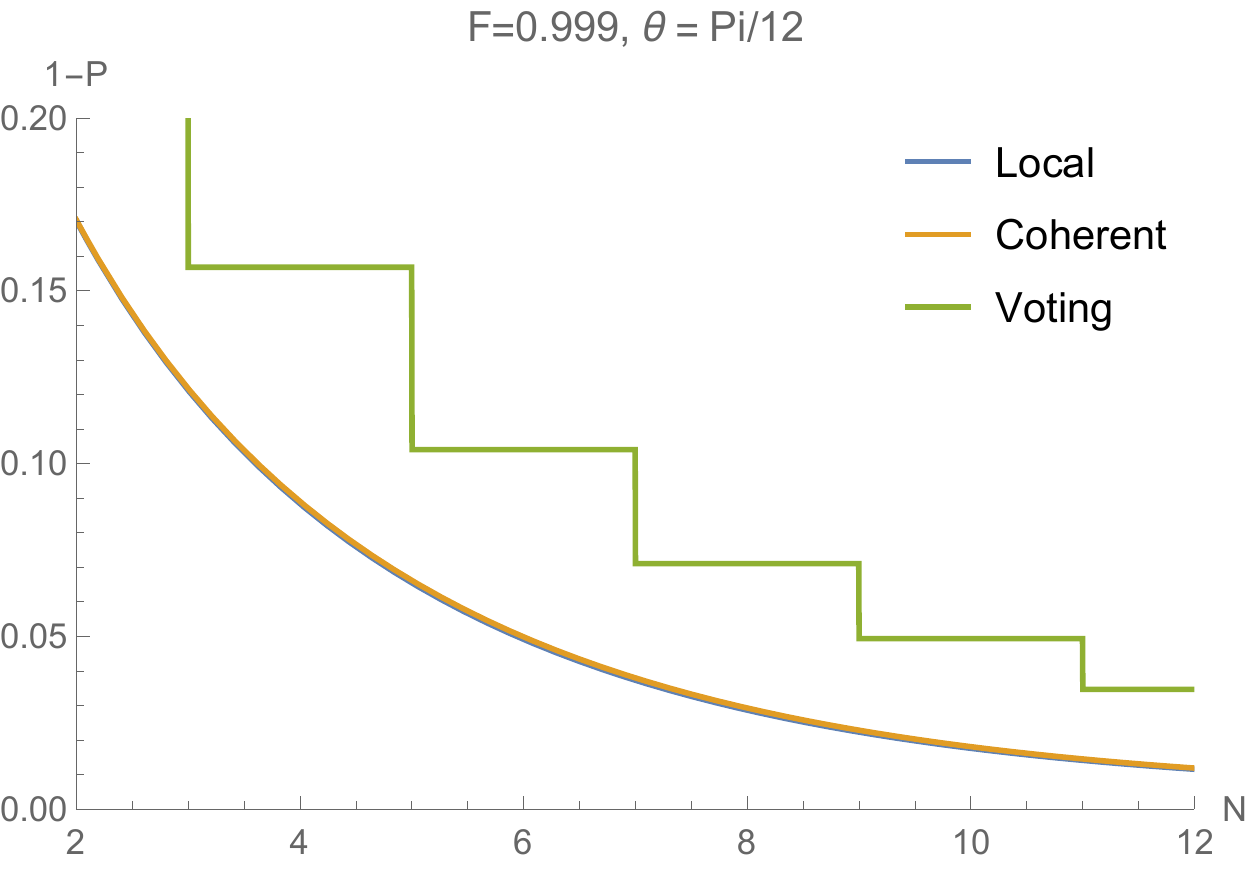} \\
  \end{tabular}
  
  \caption{The probability of failure for three different multiple-copy state discrimination schemes with imperfect preparation: local-adaptive measurements (Local, blue), quantum data gathering (Coherent, orange) and voting based on fixed measurements (Voting, green). The former two schemes do not take into account the preparation noise however the latter scheme does. A range of parameters for the angle $\theta$ and fidelity $F$ are used.  \label{plots}}
\end{figure*}

In Fig. \ref{plots} we plot, as a function of the number $N$ of resource qubits, the probability of failure for both local-adaptive measurements and quantum data gathering, alongside a majority voting fixed measurement scheme, for two values of the angle $\theta$ and the fidelity $F$. For now we focus on the former two schemes. In both cases, we have used equiprobable preparation $p_0=p_1=1/2$. Despite the range of values, some broad features emerge. We comment on the many-copy limit, in which both quantitites converge upon the same value, below. What is relevant at this point is that, in all cases, the local scheme approaches this limit with fewer qubits than the collective scheme. This improvement is small enough, in the fourth or fifth decimal place for some cases, that it is probably not experimentally significant. Nonetheless, we have shown that one property, resilience to noise, is improved by measuring locally.
\par
The third scheme plotted in Fig. \ref{plots} is a majority voting scheme in which the Helstrom measurement is performed on each qubit and the most common outcome in the measurement record is the overall outcome. There is no simple analytic expression for the success probability but it is straightforward to find numerically \cite{higginsetal2}. The Helstrom measurement is that for discriminating the two mixed states, rather than the original pure states, though this will be the same for equal priors. Thus, this measurement scheme takes into account both the whole measurement record and the noise in the preparation. In general, we find that this simple generalisation is enough to outperform the other two schemes. In particular, it is not limited by the same asymptotic behaviour as those schemes. This turns out not to be true if there is only a small amount of noise, as can be seen in the graph with $F=0.999$ and $\theta=\pi/12$. The majority voting scheme will not reach the multiple-copy Helstrom bound in any case. As the fidelity becomes closer to one, the two previously analysed schemes will become closer to the genuine optimal scheme, hence they perform better for moderate $N$ in the high-fidelity case.
\par
Special attention should be paid to the many-copy limit of both schemes. Interestingly, one finds the same value in both cases:
\begin{align}
\lim_{N \rightarrow \infty} {\rm P}^{qdg}_N &= \lim_{N \rightarrow \infty} {\rm P}^{ad}_N \nonumber \\
&= 1 - \frac{1-F}{1- (2F-1)\cos^2 ( 2 \theta) }. \label{infinitylimit}
\end{align}
In the limit $F=1$, this equation reaches unity and so the states can be perfectly discriminated given an infinite number of copies. If instead the two states are the same $\theta=0$, then we find a probability of one-half. This makes sense as it should be impossible to distinguish two equal states and all that can be done is to guess. These two limits are non-commuting. This occurs because the measurement schemes are ill-defined when discriminating equal states, i.e., the unitary operation for quantum data gathering would need to map two orthogonal states onto the same state, which is clearly not possible.
\par
That both schemes reach the same many-copy limit, which in general is less than unity, is intriguing. It suggests that there is a systematic error which arises when applying a state discrimination scheme to the wrong pair of states, which cannot be overcome by increasing the number of resource qubits.
\par
The specific form of the many-copy limit can be calculated in a different manner, by understanding the behaviour of the local-adaptive measurement scheme in such a regime. Analysing this behaviour also helps in improving intuition of that scheme. Inspection of Eq. \ref{locadopt} reveals that the scheme in this case can be understood as hypothesis checking. If the outcome on one qubit suggests that $\ket{\psi_0}$ was the prepared state, the next measurement will be onto the basis $\ket{\psi_0}, \ket{\psi_{0 \perp}}$, with the latter outcome associated with a preparation of $\ket{\psi_1}$. This explains why the strategy cannot perfectly discriminate. When applied to mixed states, neither measurement outcome is impossible. This hypothesis-checking scheme can be used to calculate the probability of success. We assume that the strategy of hypothesis checking is used for an infinite number of qubits. We find agreement with the original calculation. Two probabilities are required. Firstly,
\begin{equation}
{\rm P}(a | i_{N-1} =a , a ) = F
\end{equation}
is the probability of finding outcome $a$, given that the previous outcome was $a$ (so that the measurement at this stage is $\ket{\psi_a}, \ket{\psi_{a \perp}}$), given that $\ket{\psi_a}$ was sent. We require also
\begin{equation}
{\rm P}(a | i_{N-1} = \overline{a}, a )= F - \cos^2 ( 2 \theta) (2F-1),
\end{equation}
which is the probability of outcome $a$ (i.e., the state $\ket{\psi_{\overline{a} \perp}}$) given that the previous measurement gave the outcome $\overline{a}$, the other possible state, and that $\ket{\psi_a}$ was sent. In terms of these objects, the probability of success on the $(N+1)$th qubit is written in terms of the probability of success on the $N$th qubit:
\begin{align}
{\rm P}^{ad}_{N+1} &= {\rm P}(a | i_{N-1} =a , a ) {\rm P}^{ad}_N \\
&+ {\rm P}(a | i_{N-1} = \overline{a}, a ) (1 - {\rm P}^{ad}_N) \nonumber \\
&= F - \cos^2 ( 2 \theta) (2F-1) + \cos^2 ( 2 \theta) (2F-1) {\rm P}^{ad}_N. \nonumber 
\end{align}
This result is then used iteratively to find an expression for the probability of success after $N^{\prime}$ more measurements:
\begin{align}
{\rm P}^{ad}_{N+N^{\prime}} &= \left( \cos^2 ( 2 \theta) (2F-1) \right)^{N^{\prime}} {\rm P}^{ad}_{N}  \\
& + \left(F - \cos^2 (2 \theta) (2F-1) \right) \sum^{N^{\prime}-1}_{i=0} \cos^{2i}(2 \theta) (2F-1)^i . \nonumber
\end{align}
Finally, as $N^{\prime}$ is increased the first term will be suppressed, and in the limit of an infinite number of copies, the probability of success at a given point makes no contribution to the overall expression. All that remains is to evaluate the geometric summation and to rearrange for
\begin{equation}
\lim_{N^{\prime} \rightarrow \infty} {\rm P}^{ad}_{N+N^{\prime}} = 1 - \frac{1-F}{1- (2F-1)\cos^2(2\theta)},
\end{equation}
the same value which was found previously. Here, it has been found with a different method to the more general case. Unfortunately, a similar method for confirming the calculation does not exist for quantum data gathering, however inspection of the unitary Eq. \ref{qdgunitary} reveals similar behaviour. In the many-copy limit there, the probe states become the diagonal basis states $\ket{+},\ket{-}$. One example of the behaviour of the unitary in this regime is $U_N \ket{\psi_{0}}_{S_N} \ket{+}_A = \ket{0}_{S_N} \ket{+}_A$, so that all the unitary has done is to delete the resource qubit's information conditioned upon it matching what is already known.
\par
We have considered here only preparation noise. In the quantum data gathering scheme, there will also be noise in the gates needed to implement the unitary Eq. \ref{qdgunitary}. This operation takes the form of a rotation controlled upon binary addition of the register of each individual qubit, which can be implemented by two CNOT gates alongside single-qubit gates. Thus, $2N$ two-qubit gates are needed to perform quantum data gathering on a resource of $N$ qubits. We assume that the contribution to the noise from single qubit gates is negligible. Because the diamond norm \cite{aharanovart, aliferisetal, aharanovetal}, the standard measure of gate noise, satisfies the triangle inequality, that there are $2N$ gates required means that the total gate noise scales linearly with $N$. This will appear as a noisy channel acting upon the probe's state, and decrease further the probability of success. To make further comments, we would need to understand the form of the noise in more detail \cite{sandersdiamond}.
\par
Methods to improve the schemes exist in each case. For the local measurement scheme, we have neglected the entire measurement record and made a decision based only on the final measurement made. As discussed above, the local adaptive scheme converges to a measurement which, in the pure state case is in a basis along and orthogonal to the prepared state (the ``fully biased" measurement \cite{higginsetal}). Thus in the limit, we always obtain the outcome verifying the current guess, and the probability of error approaches zero exponentially. This procedure, however, is not robust to noise: as soon as there is any finite probability of error in measurement, whether due to imperfect measurement or noisy preparation, the Markovian scheme results in a non-zero probability of error, which is simply the probability of error in any given round. This can be improved by referring back to the whole measurement record. If we have measured $N$ systems, and $N-1$ gave the result $1$, while the $N$th gave the result $0$, we could reasonably infer that an error occured in the $N$th round, and return outcome $1$. Majority vote on the measurement record thus may be expected to recover an exponential decay in the probability of error. Note however that, although the probability of error at any given $N$ is given by Eq. \ref{infinitylimit} in the large $N$ limit, the measurement performed at each step depends on the outcome of the previous measurement. If each measurement were the same, the measurement record would be a sequence of independent and identically distributed (i.i.d.) random variables, and we could use classical statistical techniques, specifically the classical Chernoff bound \cite{higginsetal, coverandthomas} to argue that the probability of error indeed decays exponentially. Indeed, this is verified by the numerically evaluated function in Fig. \ref{plots}. As the measurement record is not i.i.d., this is not possible, and more sophisticated techniques are needed. However, we know that in this limit the measurement switches between the two possible fully biased measurements, and an upper bound on the probability of error is given by considering just the measurement outcomes for that measurement which occurs more often. This tells us that the probability of error decays with an error exponent given by the classical Chernoff bound for this measurement, and with the exponent modified by the fraction of the total measurement record considered. Further, previous work \cite{higginsetal} shows that for very high fidelities, the fixed measurement giving the best error exponent in the asymptotic limit is close to fully biased. For fidelities less than around $99\%$, the optimal fixed measurement becomes unbiased \cite{higginsetal}.
\par
In quantum data gathering, the scheme can be modified by measuring the resource qubits at each step and acting based upon the outcome. As noted, an outcome of $\ket{1}$ indicates that all quantum information gathered up until that point has been lost. So, one way to modify the protocol is to restart whenever such an outcome occurs. Some care must be taken as only a finite number of consecutive $\ket{0}$ outcomes can occur before a bad outcome. In Fig. \ref{plots} it is seen that only small number, three or four, of interactions are required to get very close to the best-possible probability. However, numerical evaluation of the relevant probabilities reveals that even this small number is unlikely enough (while still being highly probably, $p \geq 0.97$ typically) to bring the overall probability of state discrimination below that which occurs if the qubits are not measured. This is played off against two things. Firstly, success here is heralded at the expense of increasing ambiguity in some cases, similar to unambiguous state discrimination. Secondly, if there are many resource qubits available, a small run of successes becomes likely to occur at some point. Thus, in some scenarios it may be advantageous to post-select based on the measurement outcomes. A hybrid scheme in which subsets of systems are measured collectively, followed by majority voting on the measurement output would give an improved probability of success, but still less than the local scheme. Here we chose to evaluate the performance of a scheme requiring a single qubit of memory as the quantum probe. A fully general scheme, which would achieve the optimal Helstrom measurement for arbitrary many-copy states, would require a processor of size $\log N$ \cite{quantumdatagathering}. Our results show that how a collective measurement is implemented has a considerable effect on its robustness to noise.

\section{Conclusion}

We have considered the ability of two multiple-copy state discrimination schemes to perform when the state preparation is imperfect. We find two surprising results. Firstly, that for small amounts of uncharacterised noise, the optimal local adaptive measurement is more robust than the simple, single qubit collective scheme. With a simple modification of the scheme to take into account the whole measurement record, we obtain a protocol which performs better than fixed, unbiased local measurements and majority vote for small $N$, and retains the desirable exponential decay of error probability in the limit of large $N$, without requiring prior knowledge of the amount of noise. We also find that both schemes have the same many-copy limit, which is less than unity. Despite the different physical mechanisms used in each scheme, they have precisely the same behaviour in this regime. This suggests that the quantity found is a generic property of applying an incorrect scheme, and should be investigated further.
\par
It would be useful to know an optimal state discrimination scheme for mixed states of the type considered here. A natural starting point would be to generalise the local-adaptive scheme to use the entire measurement record when updating the prior probabilities in a Bayesian manner and to calculate the range, if any, in which this strategy is optimal. In general, more analytic work is required in multiple-copy state discrimination. Some of the techniques used here may be found to be useful in that task.

\acknowledgements
This work was supported by the UK Engineering and Physical Sciences Research Council
and by the Royal Society (RP150122).

\end{document}